\pdfoutput=1
\documentclass[11pt,twoside,a4paper,cmspaper,final,collab]{cms-tdr}
\def\svnVersion{13:6308M}\def\cmsCernNo{2010-010}\def\cmsCernDate{\today}

\begin{document}\cmsNoteHeader{QCD-10-003}
%
%
%

%
%
\hyphenation{env-iron-men-tal}
\hyphenation{had-ron-i-za-tion}
\hyphenation{cal-or-i-me-ter}
\hyphenation{de-vices}
%
\RCS$Revision: 6303 $
\RCS$HeadURL: svn+ssh://alverson@svn.cern.ch/reps/tdr2/papers/QCD-10-003/trunk/QCD-10-003.tex $
\RCS$Id: QCD-10-003.tex 6303 2010-05-18 09:34:54Z slacapra $
%
%
%

\newcommand {\etal}{\mbox{et al.}\xspace} 
\newcommand {\ie}{\mbox{i.e.}\xspace}     
\newcommand {\eg}{\mbox{e.g.}\xspace}     
\newcommand {\etc}{\mbox{etc.}\xspace}     
\newcommand {\vs}{\mbox{\sl vs.}\xspace}      
\newcommand {\mdash}{\ensuremath{\mathrm{-}}} 

\newcommand {\Lone}{Level-1\xspace} 
\newcommand {\Ltwo}{Level-2\xspace}
\newcommand {\Lthree}{Level-3\xspace}

\providecommand{\ACERMC} {\textsc{AcerMC}\xspace}
\providecommand{\ALPGEN} {{\textsc{alpgen}}\xspace}
\providecommand{\CHARYBDIS} {{\textsc{charybdis}}\xspace}
\providecommand{\CMKIN} {\textsc{cmkin}\xspace}
\providecommand{\CMSIM} {{\textsc{cmsim}}\xspace}
\providecommand{\CMSSW} {{\textsc{cmssw}}\xspace}
\providecommand{\COBRA} {{\textsc{cobra}}\xspace}
\providecommand{\COCOA} {{\textsc{cocoa}}\xspace}
\providecommand{\COMPHEP} {\textsc{CompHEP}\xspace}
\providecommand{\EVTGEN} {{\textsc{evtgen}}\xspace}
\providecommand{\FAMOS} {{\textsc{famos}}\xspace}
\providecommand{\GARCON} {\textsc{garcon}\xspace}
\providecommand{\GARFIELD} {{\textsc{garfield}}\xspace}
\providecommand{\GEANE} {{\textsc{geane}}\xspace}
\providecommand{\GEANTfour} {{\textsc{geant4}}\xspace}
\providecommand{\GEANTthree} {{\textsc{geant3}}\xspace}
\providecommand{\GEANT} {{\textsc{geant}}\xspace}
\providecommand{\HDECAY} {\textsc{hdecay}\xspace}
\providecommand{\HERWIG} {{\textsc{herwig}}\xspace}
\providecommand{\HIGLU} {{\textsc{higlu}}\xspace}
\providecommand{\HIJING} {{\textsc{hijing}}\xspace}
\providecommand{\IGUANA} {\textsc{iguana}\xspace}
\providecommand{\ISAJET} {{\textsc{isajet}}\xspace}
\providecommand{\ISAPYTHIA} {{\textsc{isapythia}}\xspace}
\providecommand{\ISASUGRA} {{\textsc{isasugra}}\xspace}
\providecommand{\ISASUSY} {{\textsc{isasusy}}\xspace}
\providecommand{\ISAWIG} {{\textsc{isawig}}\xspace}
\providecommand{\MADGRAPH} {\textsc{MadGraph}\xspace}
\providecommand{\MCATNLO} {\textsc{mc@nlo}\xspace}
\providecommand{\MCFM} {\textsc{mcfm}\xspace}
\providecommand{\MILLEPEDE} {{\textsc{millepede}}\xspace}
\providecommand{\ORCA} {{\textsc{orca}}\xspace}
\providecommand{\OSCAR} {{\textsc{oscar}}\xspace}
\providecommand{\PHOTOS} {\textsc{photos}\xspace}
\providecommand{\PROSPINO} {\textsc{prospino}\xspace}
\providecommand{\PYTHIA} {{\textsc{pythia}}\xspace}
\providecommand{\SHERPA} {{\textsc{sherpa}}\xspace}
\providecommand{\TAUOLA} {\textsc{tauola}\xspace}
\providecommand{\TOPREX} {\textsc{TopReX}\xspace}
\providecommand{\XDAQ} {{\textsc{xdaq}}\xspace}

\newcommand {\DZERO}{D\O\xspace}     


\newcommand{\de}{\ensuremath{^\circ}}
\newcommand{\ten}[1]{\ensuremath{\times \text{10}^\text{#1}}}
\newcommand{\unit}[1]{\ensuremath{\text{\,#1}}\xspace}
\newcommand{\mum}{\ensuremath{\,\mu\text{m}}\xspace}
\newcommand{\micron}{\ensuremath{\,\mu\text{m}}\xspace}
\newcommand{\cm}{\ensuremath{\,\text{cm}}\xspace}
\newcommand{\mm}{\ensuremath{\,\text{mm}}\xspace}
\newcommand{\mus}{\ensuremath{\,\mu\text{s}}\xspace}
\newcommand{\keV}{\ensuremath{\,\text{ke\hspace{-.08em}V}}\xspace}
\newcommand{\MeV}{\ensuremath{\,\text{Me\hspace{-.08em}V}}\xspace}
\newcommand{\GeV}{\ensuremath{\,\text{Ge\hspace{-.08em}V}}\xspace}
\newcommand{\TeV}{\ensuremath{\,\text{Te\hspace{-.08em}V}}\xspace}
\newcommand{\PeV}{\ensuremath{\,\text{Pe\hspace{-.08em}V}}\xspace}
\newcommand{\keVc}{\ensuremath{{\,\text{ke\hspace{-.08em}V\hspace{-0.16em}/\hspace{-0.08em}}c}}\xspace}
\newcommand{\MeVc}{\ensuremath{{\,\text{Me\hspace{-.08em}V\hspace{-0.16em}/\hspace{-0.08em}}c}}\xspace}
\newcommand{\GeVc}{\ensuremath{{\,\text{Ge\hspace{-.08em}V\hspace{-0.16em}/\hspace{-0.08em}}c}}\xspace}
\newcommand{\TeVc}{\ensuremath{{\,\text{Te\hspace{-.08em}V\hspace{-0.16em}/\hspace{-0.08em}}c}}\xspace}
\newcommand{\keVcc}{\ensuremath{{\,\text{ke\hspace{-.08em}V\hspace{-0.16em}/\hspace{-0.08em}}c^\text{2}}}\xspace}
\newcommand{\MeVcc}{\ensuremath{{\,\text{Me\hspace{-.08em}V\hspace{-0.16em}/\hspace{-0.08em}}c^\text{2}}}\xspace}
\newcommand{\GeVcc}{\ensuremath{{\,\text{Ge\hspace{-.08em}V\hspace{-0.16em}/\hspace{-0.08em}}c^\text{2}}}\xspace}
\newcommand{\TeVcc}{\ensuremath{{\,\text{Te\hspace{-.08em}V\hspace{-0.16em}/\hspace{-0.08em}}c^\text{2}}}\xspace}

\newcommand{\pbinv} {\mbox{\ensuremath{\,\text{pb}^\text{$-$1}}}\xspace}
\newcommand{\fbinv} {\mbox{\ensuremath{\,\text{fb}^\text{$-$1}}}\xspace}
\newcommand{\nbinv} {\mbox{\ensuremath{\,\text{nb}^\text{$-$1}}}\xspace}
\newcommand{\percms}{\ensuremath{\,\text{cm}^\text{$-$2}\,\text{s}^\text{$-$1}}\xspace}
\newcommand{\lumi}{\ensuremath{\mathcal{L}}\xspace}
\newcommand{\Lumi}{\ensuremath{\mathcal{L}}\xspace}
%
\newcommand{\LvLow}  {\ensuremath{\mathcal{L}=\text{10}^\text{32}\,\text{cm}^\text{$-$2}\,\text{s}^\text{$-$1}}\xspace}
\newcommand{\LLow}   {\ensuremath{\mathcal{L}=\text{10}^\text{33}\,\text{cm}^\text{$-$2}\,\text{s}^\text{$-$1}}\xspace}
\newcommand{\lowlumi}{\ensuremath{\mathcal{L}=\text{2}\times \text{10}^\text{33}\,\text{cm}^\text{$-$2}\,\text{s}^\text{$-$1}}\xspace}
\newcommand{\LMed}   {\ensuremath{\mathcal{L}=\text{2}\times \text{10}^\text{33}\,\text{cm}^\text{$-$2}\,\text{s}^\text{$-$1}}\xspace}
\newcommand{\LHigh}  {\ensuremath{\mathcal{L}=\text{10}^\text{34}\,\text{cm}^\text{-2}\,\text{s}^\text{$-$1}}\xspace}
\newcommand{\hilumi} {\ensuremath{\mathcal{L}=\text{10}^\text{34}\,\text{cm}^\text{-2}\,\text{s}^\text{$-$1}}\xspace}


\newcommand{\zp}{\ensuremath{\mathrm{Z}^\prime}\xspace}


\newcommand{\kt}{\ensuremath{k_{\mathrm{T}}}\xspace}
\newcommand{\BC}{\ensuremath{{B_{\mathrm{c}}}}\xspace}
\newcommand{\bbarc}{\ensuremath{{\overline{\mathrm{b}}\mathrm{c}}}\xspace}
\newcommand{\bbbar}{\ensuremath{{\mathrm{b}\overline{\mathrm{b}}}}\xspace}
\newcommand{\ccbar}{\ensuremath{{\mathrm{c}\overline{\mathrm{c}}}}\xspace}
\newcommand{\JPsi}{\ensuremath{{\mathrm{J}}\hspace{-.08em}/\hspace{-.14em}\psi}\xspace}
\newcommand{\bspsiphi}{\ensuremath{\mathrm{B}_\mathrm{s} \to \JPsi\, \phi}\xspace}
\newcommand{\AFB}{\ensuremath{A_\text{FB}}\xspace}
\newcommand{\EE}{\ensuremath{\mathrm{e}^+\mathrm{e}^-}\xspace}
\newcommand{\MM}{\ensuremath{\mu^+\mu^-}\xspace}
\newcommand{\TT}{\ensuremath{\tau^+\tau^-}\xspace}
\newcommand{\wangle}{\ensuremath{\sin^{2}\theta_{\text{eff}}^\text{lept}(M^2_\mathrm{Z})}\xspace}
\newcommand{\ttbar}{\ensuremath{{\mathrm{t}\overline{\mathrm{t}}}}\xspace}
\newcommand{\stat}{\ensuremath{\,\text{(stat.)}}\xspace}
\newcommand{\syst}{\ensuremath{\,\text{(syst.)}}\xspace}

\newcommand{\HGG}{\ensuremath{\mathrm{H}\to\gamma\gamma}}
\newcommand{\gev}{\GeV}
\newcommand{\GAMJET}{\ensuremath{\gamma + \text{jet}}}
\newcommand{\PPTOJETS}{\ensuremath{\mathrm{pp}\to\text{jets}}}
\newcommand{\PPTOGG}{\ensuremath{\mathrm{pp}\to\gamma\gamma}}
\newcommand{\PPTOGAMJET}{\ensuremath{\mathrm{pp}\to\gamma +
\mathrm{jet}
}}
\newcommand{\MH}{\ensuremath{\mathrm{M_{\mathrm{H}}}}}
\newcommand{\RNINE}{\ensuremath{\mathrm{R}_\mathrm{9}}}
\newcommand{\DR}{\ensuremath{\Delta\mathrm{R}}}


\newcommand{\PT}{\ensuremath{p_{\mathrm{T}}}\xspace}
\newcommand{\pt}{\ensuremath{p_{\mathrm{T}}}\xspace}
\newcommand{\ET}{\ensuremath{E_{\mathrm{T}}}\xspace}
\newcommand{\HT}{\ensuremath{H_{\mathrm{T}}}\xspace}
\newcommand{\et}{\ensuremath{E_{\mathrm{T}}}\xspace}
\newcommand{\Em}{\ensuremath{E\!\!\!/}\xspace}
\newcommand{\Pm}{\ensuremath{p\!\!\!/}\xspace}
\newcommand{\PTm}{\ensuremath{{p\!\!\!/}_{\mathrm{T}}}\xspace}
\newcommand{\ETm}{\ensuremath{E_{\mathrm{T}}^{\text{miss}}}\xspace}
\newcommand{\MET}{\ensuremath{E_{\mathrm{T}}^{\text{miss}}}\xspace}
\newcommand{\ETmiss}{\ensuremath{E_{\mathrm{T}}^{\text{miss}}}\xspace}
\newcommand{\VEtmiss}{\ensuremath{{\vec E}_{\mathrm{T}}^{\text{miss}}}\xspace}

%

\newcommand{\ga}{\ensuremath{\gtrsim}}
\newcommand{\la}{\ensuremath{\lesssim}}
\newcommand{\swsq}{\ensuremath{\sin^2\theta_\mathrm{W}}\xspace}
\newcommand{\cwsq}{\ensuremath{\cos^2\theta_\mathrm{W}}\xspace}
\newcommand{\tanb}{\ensuremath{\tan\beta}\xspace}
\newcommand{\tanbsq}{\ensuremath{\tan^{2}\beta}\xspace}
\newcommand{\sidb}{\ensuremath{\sin 2\beta}\xspace}
\newcommand{\alpS}{\ensuremath{\alpha_S}\xspace}
\newcommand{\alpt}{\ensuremath{\tilde{\alpha}}\xspace}

\newcommand{\QL}{\ensuremath{Q_L}\xspace}
\newcommand{\sQ}{\ensuremath{\tilde{Q}}\xspace}
\newcommand{\sQL}{\ensuremath{\tilde{Q}_L}\xspace}
\newcommand{\ULC}{\ensuremath{U_L^C}\xspace}
\newcommand{\sUC}{\ensuremath{\tilde{U}^C}\xspace}
\newcommand{\sULC}{\ensuremath{\tilde{U}_L^C}\xspace}
\newcommand{\DLC}{\ensuremath{D_L^C}\xspace}
\newcommand{\sDC}{\ensuremath{\tilde{D}^C}\xspace}
\newcommand{\sDLC}{\ensuremath{\tilde{D}_L^C}\xspace}
\newcommand{\LL}{\ensuremath{L_L}\xspace}
\newcommand{\sL}{\ensuremath{\tilde{L}}\xspace}
\newcommand{\sLL}{\ensuremath{\tilde{L}_L}\xspace}
\newcommand{\ELC}{\ensuremath{E_L^C}\xspace}
\newcommand{\sEC}{\ensuremath{\tilde{E}^C}\xspace}
\newcommand{\sELC}{\ensuremath{\tilde{E}_L^C}\xspace}
\newcommand{\sEL}{\ensuremath{\tilde{E}_L}\xspace}
\newcommand{\sER}{\ensuremath{\tilde{E}_R}\xspace}
\newcommand{\sFer}{\ensuremath{\tilde{f}}\xspace}
\newcommand{\sQua}{\ensuremath{\tilde{q}}\xspace}
\newcommand{\sUp}{\ensuremath{\tilde{u}}\xspace}
\newcommand{\suL}{\ensuremath{\tilde{u}_L}\xspace}
\newcommand{\suR}{\ensuremath{\tilde{u}_R}\xspace}
\newcommand{\sDw}{\ensuremath{\tilde{d}}\xspace}
\newcommand{\sdL}{\ensuremath{\tilde{d}_L}\xspace}
\newcommand{\sdR}{\ensuremath{\tilde{d}_R}\xspace}
\newcommand{\sTop}{\ensuremath{\tilde{t}}\xspace}
\newcommand{\stL}{\ensuremath{\tilde{t}_L}\xspace}
\newcommand{\stR}{\ensuremath{\tilde{t}_R}\xspace}
\newcommand{\stone}{\ensuremath{\tilde{t}_1}\xspace}
\newcommand{\sttwo}{\ensuremath{\tilde{t}_2}\xspace}
\newcommand{\sBot}{\ensuremath{\tilde{b}}\xspace}
\newcommand{\sbL}{\ensuremath{\tilde{b}_L}\xspace}
\newcommand{\sbR}{\ensuremath{\tilde{b}_R}\xspace}
\newcommand{\sbone}{\ensuremath{\tilde{b}_1}\xspace}
\newcommand{\sbtwo}{\ensuremath{\tilde{b}_2}\xspace}
\newcommand{\sLep}{\ensuremath{\tilde{l}}\xspace}
\newcommand{\sLepC}{\ensuremath{\tilde{l}^\mathrm{C}}\xspace}
\newcommand{\sEl}{\ensuremath{\tilde{\mathrm{e}}}\xspace}
\newcommand{\sElC}{\ensuremath{\tilde{\mathrm{e}}^\mathrm{C}}\xspace}
\newcommand{\seL}{\ensuremath{\tilde{\mathrm{e}}_\mathrm{L}}\xspace}
\newcommand{\seR}{\ensuremath{\tilde{\mathrm{e}}_\mathrm{R}}\xspace}
\newcommand{\snL}{\ensuremath{\tilde{\nu}_L}\xspace}
\newcommand{\sMu}{\ensuremath{\tilde{\mu}}\xspace}
\newcommand{\sNu}{\ensuremath{\tilde{\nu}}\xspace}
\newcommand{\sTau}{\ensuremath{\tilde{\tau}}\xspace}
\newcommand{\Glu}{\ensuremath{g}\xspace}
\newcommand{\sGlu}{\ensuremath{\tilde{g}}\xspace}
\newcommand{\Wpm}{\ensuremath{\mathrm{W}^{\pm}}\xspace}
\newcommand{\sWpm}{\ensuremath{\tilde{\mathrm{W}}^{\pm}}\xspace}
\newcommand{\Wz}{\ensuremath{\mathrm{W}^{0}}\xspace}
\newcommand{\sWz}{\ensuremath{\tilde{\mathrm{W}}^{0}}\xspace}
\newcommand{\sWino}{\ensuremath{\tilde{\mathrm{W}}}\xspace}
\newcommand{\Bz}{\ensuremath{\mathrm{B}^{0}}\xspace}
\newcommand{\sBz}{\ensuremath{\tilde{\mathrm{B}}^{0}}\xspace}
\newcommand{\sBino}{\ensuremath{\tilde{\mathrm{B}}}\xspace}
\newcommand{\Zz}{\ensuremath{\mathrm{Z}^{0}}\xspace}
\newcommand{\sZino}{\ensuremath{\tilde{\mathrm{Z}}^{0}}\xspace}
\newcommand{\sGam}{\ensuremath{\tilde{\gamma}}\xspace}
\newcommand{\chiz}{\ensuremath{\tilde{\chi}^{0}}\xspace}
\newcommand{\chip}{\ensuremath{\tilde{\chi}^{+}}\xspace}
\newcommand{\chim}{\ensuremath{\tilde{\chi}^{-}}\xspace}
\newcommand{\chipm}{\ensuremath{\tilde{\chi}^{\pm}}\xspace}
\newcommand{\Hone}{\ensuremath{\mathrm{H}_\mathrm{d}}\xspace}
\newcommand{\sHone}{\ensuremath{\tilde{\mathrm{H}}_\mathrm{d}}\xspace}
\newcommand{\Htwo}{\ensuremath{\mathrm{H}_\mathrm{u}}\xspace}
\newcommand{\sHtwo}{\ensuremath{\tilde{\mathrm{H}}_\mathrm{u}}\xspace}
\newcommand{\sHig}{\ensuremath{\tilde{\mathrm{H}}}\xspace}
\newcommand{\sHa}{\ensuremath{\tilde{\mathrm{H}}_\mathrm{a}}\xspace}
\newcommand{\sHb}{\ensuremath{\tilde{\mathrm{H}}_\mathrm{b}}\xspace}
\newcommand{\sHpm}{\ensuremath{\tilde{\mathrm{H}}^{\pm}}\xspace}
\newcommand{\hz}{\ensuremath{\mathrm{h}^{0}}\xspace}
\newcommand{\Hz}{\ensuremath{\mathrm{H}^{0}}\xspace}
\newcommand{\Az}{\ensuremath{\mathrm{A}^{0}}\xspace}
\newcommand{\Hpm}{\ensuremath{\mathrm{H}^{\pm}}\xspace}
\newcommand{\sGra}{\ensuremath{\tilde{\mathrm{G}}}\xspace}
\newcommand{\mtil}{\ensuremath{\tilde{m}}\xspace}
\newcommand{\rpv}{\ensuremath{\rlap{\kern.2em/}R}\xspace}
\newcommand{\LLE}{\ensuremath{LL\bar{E}}\xspace}
\newcommand{\LQD}{\ensuremath{LQ\bar{D}}\xspace}
\newcommand{\UDD}{\ensuremath{\overline{UDD}}\xspace}
\newcommand{\Lam}{\ensuremath{\lambda}\xspace}
\newcommand{\Lamp}{\ensuremath{\lambda'}\xspace}
\newcommand{\Lampp}{\ensuremath{\lambda''}\xspace}
\newcommand{\spinbd}[2]{\ensuremath{\bar{#1}_{\dot{#2}}}\xspace}

\newcommand{\MD}{\ensuremath{{M_\mathrm{D}}}\xspace}
\newcommand{\Mpl}{\ensuremath{{M_\mathrm{Pl}}}\xspace}
\newcommand{\Rinv} {\ensuremath{{R}^{-1}}\xspace}

%
%
\hyphenation{en-viron-men-tal}

\providecommand{\dedx}{\ensuremath{\text{d}E/\text{d}x}\xspace}
\providecommand{\GeVinv}{\ensuremath{\,\text{Ge\hspace{-.08em}V}^{-1}}}
\cmsNoteHeader{QCD-10-003} 
\ifthenelse{\boolean{cms@external}}{
\title{First Measurement of Bose--Einstein Correlations in proton-proton
    Collisions at
    \texorpdfstring{ \boldmath{$\sqrt{s}=0.9$} and \boldmath{$2.36\TeV$}}{sqrt(s)=0.9 and 2.36 TeV}
at the LHC}
}{
\title{First Measurement of Bose--Einstein Correlations in proton-proton
    Collisions at
\texorpdfstring{{$\sqrt{s}=$}\,0.9 and 2.36\TeV}{sqrt(s)=0.9 and 2.36 TeV}
at the LHC}
}

\date{\today}

\abstract{
    Bose--Einstein correlations  have been  measured using samples
of proton-proton collisions at 0.9 and 2.36\TeV center-of-mass energies, recorded
  by the CMS experiment at the CERN Large Hadron Collider.
    The signal is observed in the form of an enhancement of pairs of
    same-sign charged particles with small relative four-momentum.
    The size of the correlated particle emission region is seen to increase
    significantly with the particle multiplicity of the event.
}

\hypersetup{%
pdfauthor={CMS Collaboration},%
pdftitle={Measurement of Bose-Einstein correlations with first CMS data},%
pdfsubject={CMS Bose-Einstein},%
pdfkeywords={CMS, physics, QCD, low-pt, Bose-Einstein}}

\maketitle 


In particle collisions, the space-time structure of the hadronization source
can be studied using measurements of Bose--Einstein correlations (BEC) between pairs of identical bosons.
Since the first observation of BEC fifty years ago in proton-antiproton
interactions~\cite{goldhaber}, a number of measurements have been made by several experiments using different
initial states;
a detailed list of the experimental results can be found in~\cite{alex,revbec}.
Boson interferometry at the Large Hadron Collider provides a powerful
tool to investigate 
the space-time structure of the particle emission source on femtometric
length scales at
different center-of-mass
energies and with different initial states, using the same detector.
This letter reports the first measurement of BEC parameters in $pp$ collisions at 0.9 and
2.36\TeV with the CMS detector.

Constructive interference affects the joint probability for the emission of a
pair of identical bosons with four-momenta $p_1$ and $p_2$.
Experimentally, the proximity in phase space between final-state particles
is quantified by
 the Lorentz-invariant quantity
$Q =\sqrt{-(p_1-p_2)^2}=\sqrt{M^2-4m_\pi^2}$, where $M$ is the invariant mass of the two particles,
assumed to be pions with mass  $m_{\pi}$.
The BEC effect is observed as an enhancement at low $Q$ of the ratio
of the $Q$ distributions for pairs of identical particles in the same event, and
for pairs of particles in a reference sample that by construction is expected to include no BEC effect:
\begin{linenomath}
\begin{equation}
\label{eq:R1}
 R(Q) =  (\text{d}N / \text{d}Q)/ (\text{d}N_{\text{ref}} / \text{d}Q),
\end{equation}
which is then fitted with the parameterization
\begin{equation}
\label{eq:exp}
 R(Q) = C\left[1+ \lambda \Omega(Qr)\right] \cdot (1+ \delta Q).
\end{equation}
\end{linenomath}
\noindent
In a static model of particle sources, $\Omega(Qr)$ is the Fourier transform of the spatial distribution
of the emission region
of bosons with overlapping wave functions, characterized by an effective size $r$.
It is often parameterized as an exponential function,
 $\Omega(Qr)=e^{-Qr}$, or with a Gaussian form, $\Omega(Qr)=e^{-(Qr)^2}$
(\cite{kozlov0} and references therein).
The parameter $\lambda$
reflects the BEC strength for incoherent boson emission from independent sources,
$\delta$ accounts for long-range momentum correlations, and $C$ is a normalization factor.

The data used for the present analysis were collected by the CMS experiment in December 2009
from proton-proton 
collisions at center-of-mass energies of 0.9 and 2.36\TeV. A detailed description
of the CMS detector can be found in \cite{cmstdr}.
The central feature
of the CMS apparatus is a superconducting solenoid of 6 m internal diameter, providing a
uniform magnetic field of 3.8 T.
The inner tracking system is the most relevant detector for the present analysis. It is composed
of a pixel detector with three barrel layers at radii between 4.4 and 10.2\cm and a silicon
strip tracker with 10 barrel detection layers extending outwards to a radius of 1.1~m.
Each system is completed by two endcaps, extending the acceptance up to a pseudorapidity  $|\eta|= 2.5$.
The transverse-momentum ($\pt$) resolution, for 1\GeV charged particles, is between $0.7\%$ at $\eta=0$ and $2\%$ at $|\eta|=2.5$.
The events were selected by requiring activity in both beam scintillator
counters~\cite{dNdEta}.
A minimum-bias Monte Carlo (MC) sample was
generated using PYTHIA (with D6T tune)~\cite{pythia64} followed by full detector simulation based on the Geant4 program~\cite{GEANT4}.
Additional PYTHIA MC samples were generated to simulate BEC effects
with both Gaussian and  exponential forms of $\Omega(Qr)$.

Charged particles are required to have $\pt>200$\MeV, which is sufficient for
particles emitted from the interaction region
to cross
all three barrel layers of the pixel detector
and ensure good two-track separation.
Their pseudorapidity is required to satisfy $|\eta_{\text{track}}|<2.4$. 
To ensure high purity of the primary track selection, the trajectories
are required to be reconstructed in fits
with more than five degrees of freedom (dof) and $\chi^2/N_{{\rm dof}}<5.0$.
The transverse impact parameter with respect to the collision point is required
to satisfy $|d_{xy}|<0.15$~\cm. The innermost measured point of the track must
be less than $20\cm$ from the beam axis
, in order to
reduce
electrons and
positrons from photon conversions in the detector material and
secondary particles from the decay of long-lived hadrons ($K_{S}^0, \Lambda$, etc.).
In a total of 270\,472 (13\,548) events selected at 0.9 (2.36)\TeV center-of-mass energy,
2\,903\,754 (188\,140) tracks are accepted by these selection criteria.

All pairs of same-charge particles with
$Q$ between $0.02$ and $2\GeV$
are used for the measurement.
The lower limit is chosen to avoid cases of tracks that are duplicated or not well separated, while
the upper limit extends far enough beyond the signal region to verify a good match between signal
and reference samples.
A study with simulated data shows that the ratio of the tracking efficiencies of particle pairs in the signal and in
the reference samples is independent of $Q$ in the measurement region.

Coulomb interactions between charged particles modify their relative momentum distribution.
This effect, which differs for pairs with same charge (repulsion) and opposite charge (attraction),
is corrected for by using Gamow factors~\cite{noGamov}.
As a cross-check, the enhancement in the production of opposite-charge particle pairs with small values of $Q$
is measured in the data and is found to be reproduced by the Gamow factors to within $\pm 15\%$.

Different methods are designed to pair uncorrelated charged particles and
to define reference samples used to extract the distribution in the denominator of Eq.~(\ref{eq:R1}).
{\it Opposite-charge pairs}: this data set is a natural choice but contains resonances ($\eta$, $\rho$, ...)
which are not present in the same-charge combinations. 
{\it  Opposite-hemisphere pairs}: tracks are
paired after inverting in space the three-momentum of one of the
two particles: \mbox{$(E,\vec{p})\to(E,-\vec{p})$} ;
this procedure is applied to pairs with same and opposite
charges.
{\it Rotated particles}:
particle pairs are constructed after
inverting
the $x$ and $y$ components
of the three-momentum of one of the two particles: $(p_x,p_y,p_z)\to(-p_x,-p_y,p_z)$.
{\it Pairs from mixed events: }
particles from different events are combined with the
following methods: i) events are mixed at random; ii) events with similar
  charged particle multiplicity in the same $\eta$ regions are selected;
  iii) events with an invariant mass of all charged particles similar
  to that of the signal are used to form the pairs.

\begin{figure}[!h]
\ifthenelse{\boolean{cms@external}}{
    }{
    \begin{center}
}
        \includegraphics[height=0.48\textwidth,angle=90, trim = 0mm 18mm 10mm 0mm, clip]{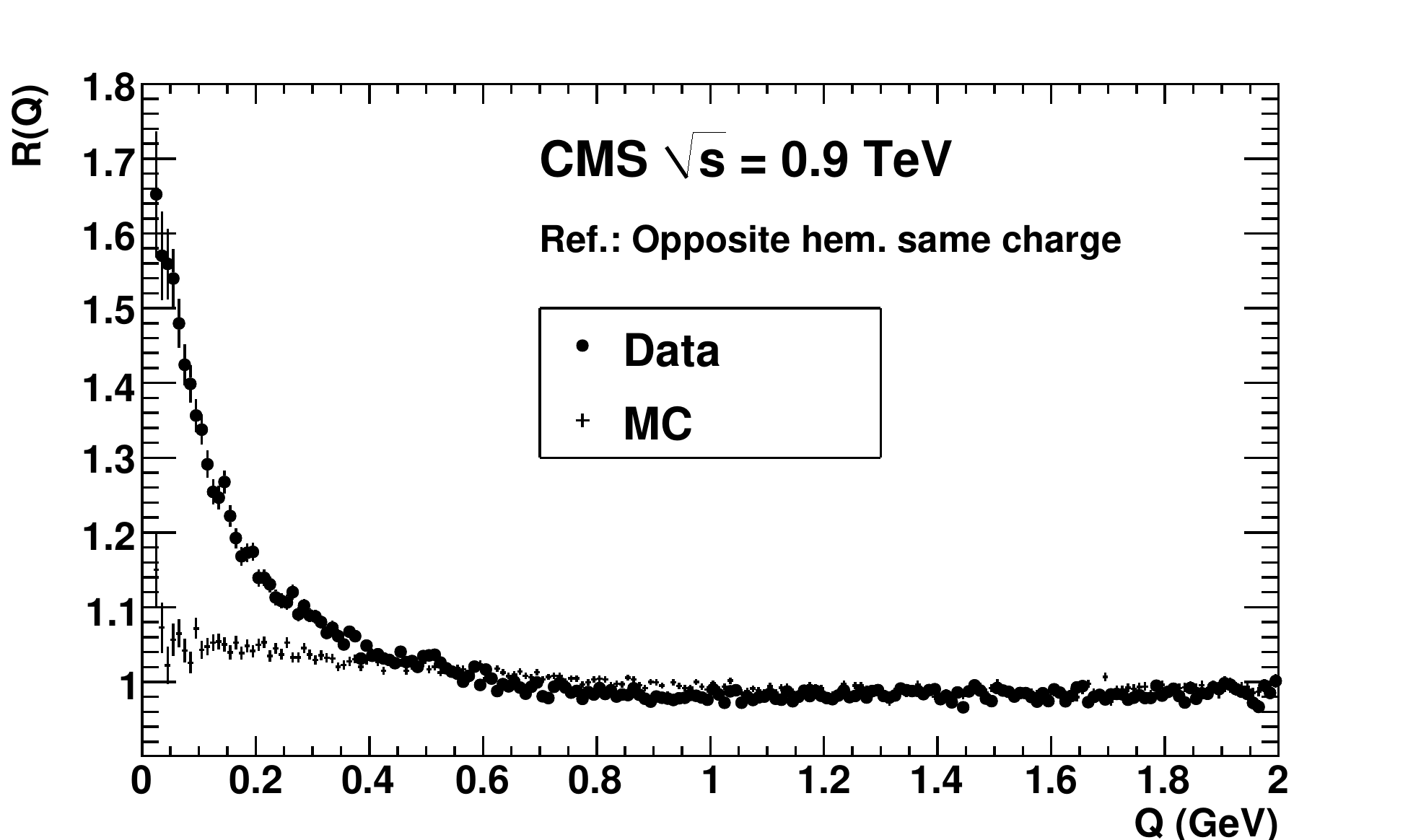}
        \caption{
	Ratios $R(Q)$ obtained with the
	opposite-hemisphere, same-charge reference samples
            for data (dots) and MC with no BEC effect (crosses).
        }
        \label{f:fig0}
\ifthenelse{\boolean{cms@external}}{
}{
    \end{center}
}
\end{figure}

As an example, the ratios $R(Q)$ obtained with the
opposite-hemisphere, same-charge reference samples are shown in Fig.~\ref{f:fig0} both for
data and simulation without BEC. A significant excess at small values of $Q$ is
observed in the data.
Additional details are given in \cite{bec_pas}.

In order to reduce the bias due to the construction of the reference samples, a double ratio $\cal{R}$ is
defined:
\begin{linenomath}
\begin{equation}
\label{eq:doubleratio}
{\cal{R}}(Q) = \frac{R}{R_{\text{MC}}} =
\left(\frac {\text{d}N/\text{d}Q} {\text{d}N_{\text{ref}}/\text{d}Q} \right)
\Big/
\left(\frac {\text{d}N_{\text{MC}}/\text{d}Q}{\text{d}N_{\text{MC,ref}}/\text{d}Q}\right),
\end{equation}
\end{linenomath}
\noindent
where the subscripts ``MC''  and ``MC,ref'' refer to the corresponding distributions from the MC simulated data
generated without BEC effects.

The results of fits of ${\cal{R}}(Q)$ based on the parameterization of Eq.~(\ref{eq:exp}) with  $\Omega(Qr)= e^{-Qr}$
are given in Table~\ref{t:resultsDR}, both for 0.9 and 2.36\TeV data.
In the case of the opposite-charge sample, it is found that the region with \mbox{$ 0.6 < Q < 0.9$\GeV},
containing a sizeable contribution of pairs from $\rho \rightarrow \pi^+ \pi^-$ decays,
is not well described by the MC \cite{bec_pas}.
This region is therefore excluded from
the fits with this reference sample and also with the combined sample defined below.

\begin{figure}[!h]
\ifthenelse{\boolean{cms@external}}{
    }{
    \begin{center}
}
        \includegraphics[height=0.48\textwidth,angle=90, trim = 0mm 18mm 10mm 0mm,clip]{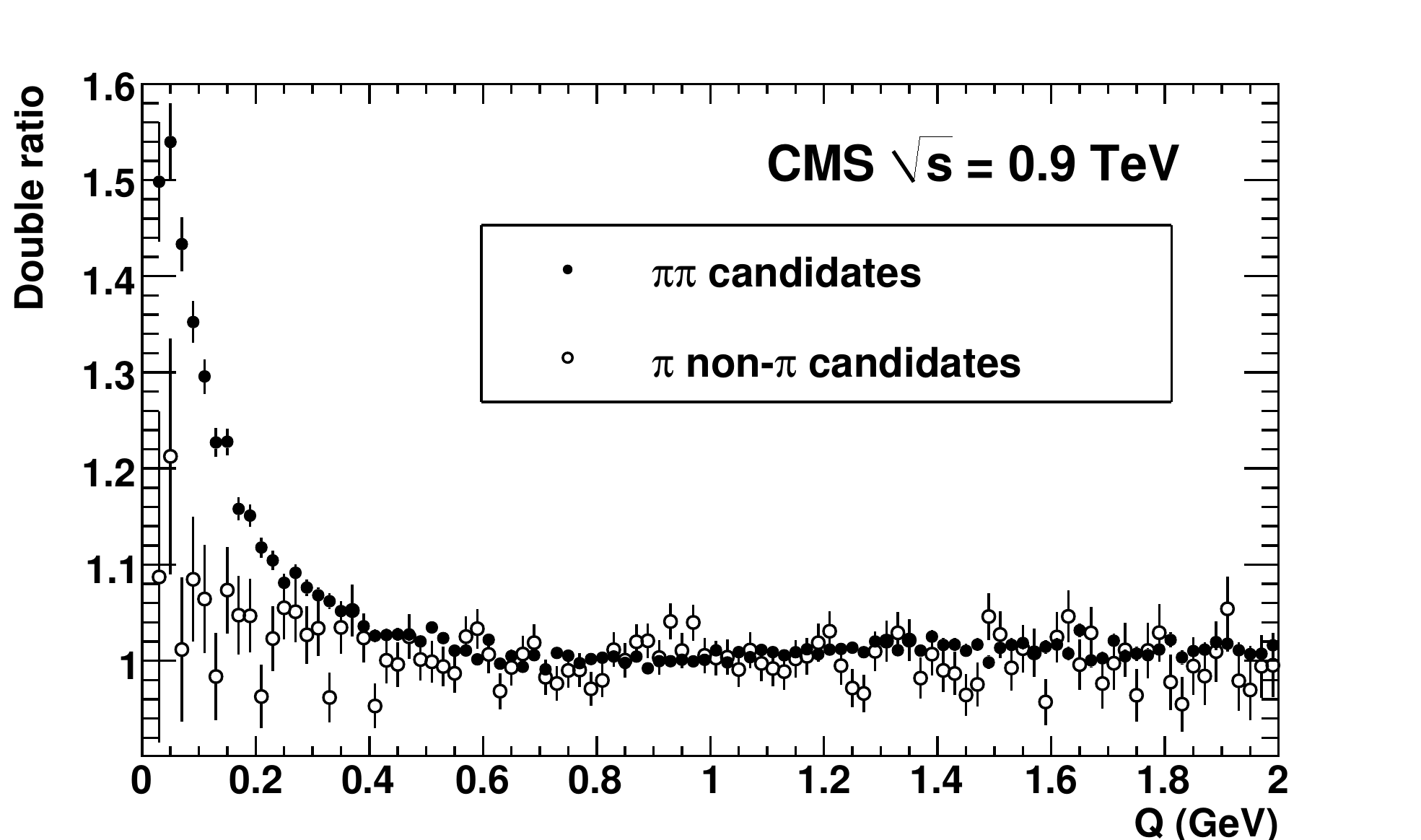}
        \caption{Double ratios ${\cal{R}}(Q)$ for the 0.9\TeV data,
using the opposite-hemisphere, same-charge reference samples
for combinations enriched, using a \dedx  measurement, in pion-pion pairs (dots)
and in pion--non-pion pairs (open circles), respectively.
}
        \label{f:resultspiNOTpi}
\ifthenelse{\boolean{cms@external}}{
}{
    \end{center}
}
\end{figure}

As a cross-check, the \dedx ~\cite{dedx}   measurements of particles in the tracker are used to
select a sample enriched in $\pi\pi$ pairs, and another sample with one of the particles not consistent with the
pion hypothesis. Figure~\ref{f:resultspiNOTpi} presents the double ratios for these two samples at  $\sqrt{s}=0.9$\TeV,
showing that an enhancement at small $Q$ values is observed only in the case of identified $\pi\pi$ pairs.

\ifthenelse{\boolean{cms@external}}{
\begin{table*}[t]
\caption{ Results of fits to the double ratios ${\cal R}(Q)$ for several reference
samples, using the parameterization of Eq.~(\ref{eq:exp}) with the exponential form,
for 0.9\TeV data (left) and 2.36\TeV data (right). Errors are statistical only, and quoted as if independent.}
\label{t:resultsDR}
{\scriptsize
\begin{tabular}{l||c|c|c|c|r||c|c|c|c|r}
\hline
\hline
&\multicolumn{4}{c}{ Results of fits to 0.9\TeV data} &&\multicolumn{5}{c}{ Results of fits to 2.36\TeV data}  \\
\hline
Reference &  $p$ value & $C$ & $\lambda$ & $r$ (fm)     & $\delta$ ($10^{-3}$  & $p$ value & $C$ & $\lambda$ & $r$ (fm)     & $\delta$ ($10^{-3}$  \\
sample     &      (\%)       &        &                   & &$\GeVinv$) & (\%) &        &                   &                   &$\GeVinv$) \\
\hline
Opposite charge               & $21.9$ & $0.988\pm 0.003$ & $0.56\pm 0.03$ & $1.46\pm0.06$ & $-4\pm 2$
  & $57$  & $1.004\pm 0.008$ & $0.53\pm 0.08$ & $1.65\pm 0.23$ & $-16\pm 6$  \\
Opposite hem. same ch.   & $7.3$   & $0.978\pm 0.003$ &$0.63\pm 0.03$ & $1.50\pm 0.06$  & $ 11\pm 2$
  & $42$  & $0.977\pm 0.006$ & $0.68\pm 0.11$ & $1.95\pm 0.24$ & $15\pm 5$  \\
Opposite hem. opp. ch.    & $11.9$ & $0.975\pm 0.003$ &$0.59\pm 0.03$ & $1.42\pm 0.06$  & $ 13\pm 2$
  & $46$  & $0.969\pm 0.005$ & $0.70\pm 0.11$ & $2.02\pm 0.23$ & $24\pm 5$  \\
Rotated                             & $0.02$ & $0.929\pm 0.003$ &$0.68\pm 0.02$ & $1.29\pm 0.04$ & $ 58\pm 3$
 & $42$  & $0.933\pm 0.007$ & $0.61\pm 0.07$ & $1.49\pm 0.15$ & $58\pm 6$  \\
Mixed evts.  (random)       & $1.9$   & $1.014\pm 0.002$ & $0.62\pm 0.04$ &$1.85\pm 0.09$ & $-20\pm 2$
 & $23$  & $1.041\pm 0.005$ & $0.74\pm 0.15$ & $2.78\pm 0.36$ & $-40\pm 4$  \\
Mixed evts. (same mult.)   & $12.2$ & $0.981\pm 0.002$ & $0.66\pm 0.03$ & $1.72\pm0.06$ & $11\pm 2$
& $35$  & $0.974\pm 0.005$ & $0.63\pm 0.10$ & $2.01\pm 0.23$ & $20\pm 5$  \\
Mixed evts.  (same mass)  & $17.0$ & $0.976\pm 0.002$ & $0.60\pm 0.03$ & $1.59\pm0.06$ & $14\pm 2$
& $73$  & $0.964\pm 0.005$ & $0.73\pm 0.11$ & $2.18\pm 0.23$ & $28\pm 5$  \\
\hline
Combined  & $2.9 $  & $0.984\pm 0.002$ & $0.63\pm 0.02$ & $1.59\pm 0.05$ & $8\pm 2$
& $89$  & $0.981\pm 0.005$ & $0.66\pm 0.07$ & $1.99\pm 0.18$ & $13\pm 4$  \\
\hline
\end{tabular}
}
\end{table*}
}{

\begin{table*}[t]
\begin{center}
\caption{ Results of fits to the double ratios ${\cal R}$ for several reference
samples, using the parameterization of Eq.~(\ref{eq:exp}) with the exponential form,
for 0.9\TeV data (top) and 2.36\TeV data (bottom). Errors are statistical only, and quoted as if independent.}
\label{t:resultsDR}
{\footnotesize
\begin{tabular}{l||c|c|c|c|c}
\hline
\multicolumn{6}{c}{ Results of fits to 0.9\TeV data} \\
\hline
Reference sample&  $p$ value (\%)& $C$ & ~$\lambda$ & $r$ (fm)     & $\delta$ ($10^{-3}\GeVinv$)  \\
\hline
Opposite charge               & $21.9$ & $0.988\pm 0.003$ & $0.56\pm 0.03$ & $1.46\pm0.06$ & $-4\pm 2$ \\
Opposite hem. same ch.   & $~~7.3$   & $0.978\pm 0.003$ &$0.63\pm 0.03$ & $1.50\pm 0.06$  & $~11\pm 2$ \\
Opposite hem. opp. ch.    & $11.9$ & $0.975\pm 0.003$ &$0.59\pm 0.03$ & $1.42\pm 0.06$  & $~13\pm 2$ \\
Rotated                             & $~~~~0.02$ & $0.929\pm 0.003$ &$0.68\pm 0.02$ & $1.29\pm 0.04$ & $~58\pm 3$ \\
Mixed evts.  (random)       & $~~1.9$   & $1.014\pm 0.002$ & $0.62\pm 0.04$ &$1.85\pm 0.09$ & $-20\pm 2~~$ \\
Mixed evts. (same mult.)   & $12.2$ & $0.981\pm 0.002$ & $0.66\pm 0.03$ & $1.72\pm0.06$ & $~11\pm 2$ \\
Mixed evts.  (same mass)  & $17.0$ & $0.976\pm 0.002$ & $0.60\pm 0.03$ & $1.59\pm0.06$ & $~14\pm 2$ \\
\hline
Combined  & $~~2.9 $  & $0.984\pm 0.002$ & $0.63\pm 0.02$ & $1.59\pm 0.05$ & $~~~8\pm 2$ \\
\hline
\end{tabular}

\vspace*{5mm}

\begin{tabular}{l||c|c|c|c|c}
\hline
\multicolumn{6}{c}{ Results of fits to 2.36\TeV data}  \\
\hline
Reference sample&  $p$ value (\%)& $C$ & $\lambda$ & $r$ (fm)     & $\delta$ ($10^{-3}\GeVinv$)  \\
\hline
Opposite charge               & $57$  & $1.004\pm 0.008$ & $0.53\pm 0.08$ & $1.65\pm 0.23$ & $-16\pm 6~\,~$  \\
Opposite hem. same ch.   & $42$  & $0.977\pm 0.006$ & $0.68\pm 0.11$ & $1.95\pm 0.24$ & $~15\pm 5$  \\
Opposite hem. opp. ch.   & $46$  & $0.969\pm 0.005$ & $0.70\pm 0.11$ & $2.02\pm 0.23$ & $~24\pm 5$  \\
Rotated                  & $42$  & $0.933\pm 0.007$ & $0.61\pm 0.07$ & $1.49\pm 0.15$ & $~58\pm 6$  \\
Mixed evts.  (random)       & $23$  & $1.041\pm 0.005$ & $0.74\pm 0.15$ & $2.78\pm 0.36$ & $-40\pm 4~\,~$  \\
Mixed evts. (same mult.)   & $35$  & $0.974\pm 0.005$ & $0.63\pm 0.10$ & $2.01\pm 0.23$ & $~20\pm 5$  \\
Mixed evts.  (same mass)  & $73$  & $0.964\pm 0.005$ & $0.73\pm 0.11$ & $2.18\pm 0.23$ & $~28\pm 5$  \\
\hline
Combined  & $89$  & $0.981\pm 0.005$ & $0.66\pm 0.07$ & $1.99\pm 0.18$ & $~13\pm 4$  \\
\hline
\end{tabular}

}
\end{center}
\end{table*}

}

As none of the definitions of the reference samples is preferable {\it a priori},
an additional, ``combined'' double ratio ${\cal{R}}^{\text{comb}}$ is formed,
where the data and MC distributions are obtained by summing the $Q$ distributions of the seven corresponding reference samples.

\begin{figure}[h!]
\ifthenelse{\boolean{cms@external}}{
    }{
    \begin{center}
}
        \includegraphics[height=0.48\textwidth, angle=90, trim = 0mm 18mm 16mm 0mm, clip]{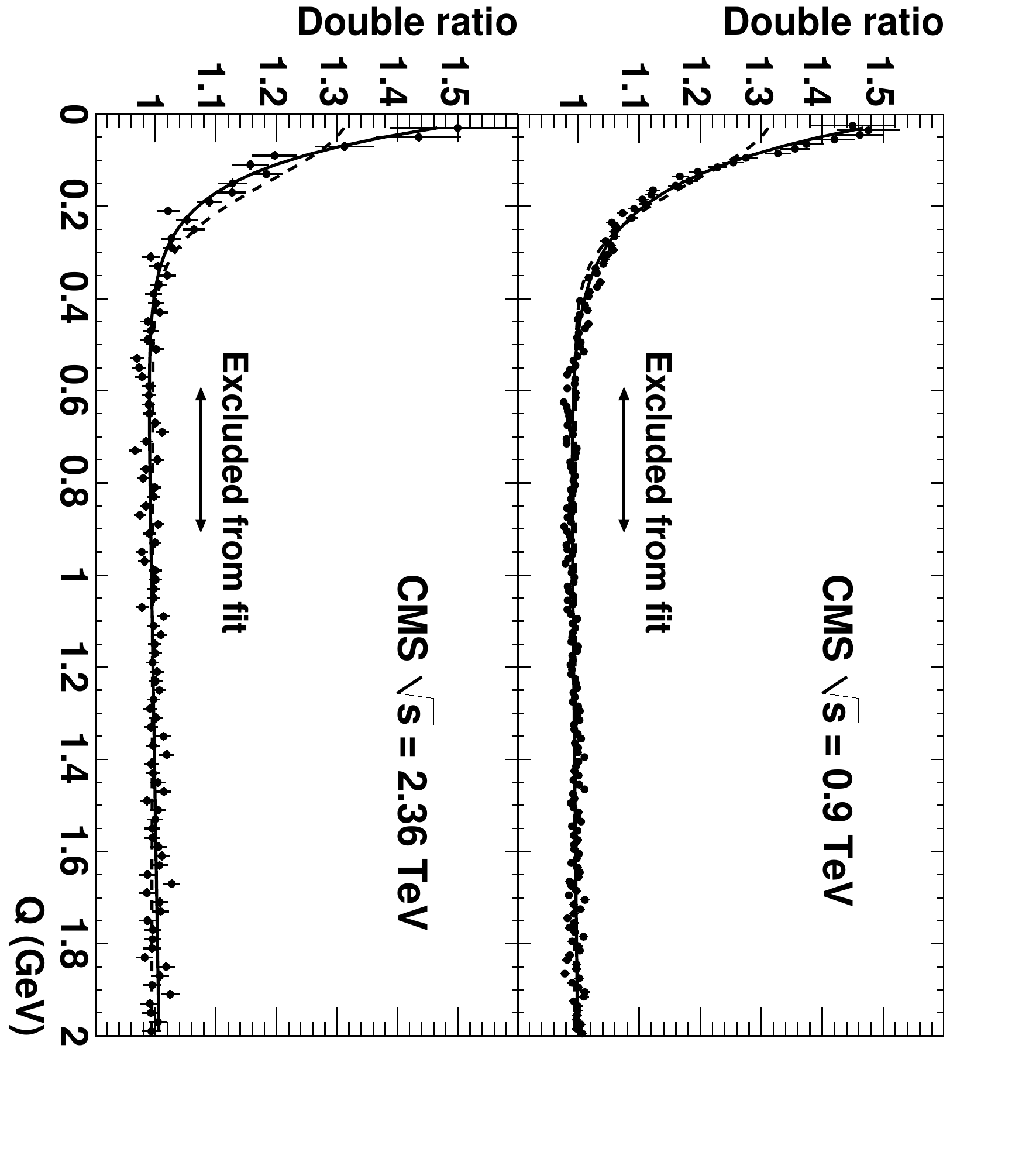}
        \caption{ Fits to the double ratios
	${\cal{R}}^{\text{comb}}(Q)$ with  exponential (solid lines) and Gaussian (dashed lines) functions,
	  for $0.9$\TeV (top) and $2.36$\TeV (bottom) data.
	  The range $0.6<Q<0.9$\GeV is excluded from the fits.}
        \label{f:combo}
\ifthenelse{\boolean{cms@external}}{
    }{
\end{center}
}
\end{figure}

The distributions of ${\cal{R}}^{\text{comb}}$ for 0.9 and 2.36\TeV data are shown in
Fig.~\ref{f:combo}, and the values of the fit parameters are given in Table~\ref{t:resultsDR}.
A large correlation is found between the parameters $\lambda$ and $r$, as well as between
$\delta$ and $C$  (correlation coefficients of $0.82$ and $-0.97$ at $0.9$\TeV, respectively).
The data are described by Eq.~(\ref{eq:exp}) with an exponential form for $\Omega(Qr)$,
as shown by the solid lines in Fig.~\ref{f:combo} and confirmed by the fit probability
($p$ value) in  Table~\ref{t:resultsDR}.
The fit with a Gaussian form, $\Omega(Qr)=~e^{-{(Qr)}^2}$,
which yields \mbox{$\lambda = 0.32\pm0.01$}, \mbox{$r =0.98\pm0.03$ fm},
does not correctly describe the ${\cal{R}}(Q)$  distribution,
as shown by the dashed lines in Fig.~\ref{f:combo} and by
a $p$ value of  $10^{-21}$.
Gaussian shape fits also proved to offer a poor description of the data in previous
measurements \cite{H1,kozlov,kittel}.

Although the values of $r$ obtained in the exponential fits cannot be compared directly
with results obtained with a Gaussian function, it should be noted
for comparison purposes that the first moment of the $\Omega(Qr)$ distribution
corresponds to $1/r$ for an exponential shape
and to $\frac{1}{r\sqrt{\pi}}$ for a Gaussian form.
Alternative functions, as defined in
\ifthenelse{\boolean{cms@external}}{\cite{kozlov,Biyajima,*BiyajimaAdd}}%
    {\cite{kozlov,Biyajima,BiyajimaAdd}}
, also describe the data well
with similar $p$ values.
In particular for the L\'evy parameterization, $\Omega(Qr)= e^{-(Qr)^\alpha}$,
the fitted values are \mbox{$\lambda = 0.93\pm0.11$}, \mbox{$r =2.46\pm0.38$ fm},
and \mbox{$\alpha =0.76\pm0.06$}, with a $p$ value of $12.8\%$.

The leading source of systematic uncertainty on the measurements arises from the fact that none of the reference samples is expected to give a perfect
description of the $Q$ distribution in the absence of BEC, and that none of them can be preferred or discarded {\it a priori}.
The corresponding contribution to the systematic error is computed as the r.m.s.\ spread
between the results obtained for the different samples, i.e., $\pm7\%$ for $\lambda$ and $\pm12\%$ for $r$.
The systematic uncertainty related to the Coulomb corrections is computed by propagating the measured $\pm 15 \%$
agreement margin, resulting in $\pm 2.8\%$ variation for $\lambda$ and $\pm 0.8\%$ for $r$.
The presence of a possible bias introduced by the track reconstruction and
selection requirements was studied by comparing
the results obtained at the generator and reconstruction levels
in the MC simulation that incorporates BEC effects.
The differences in the fitted parameter values for the different reference samples are smaller
than the statistical errors and
no systematic bias is observed for $r$. No correction is therefore applied and no additional systematic error is included.
For the $2.36$\TeV data the same relative systematic uncertainties as for the
$0.9$\TeV results are used, in view of the reduced size of the sample and the larger
statistical uncertainties of the fit results.

The BEC parameters measured with the combined reference sample are
\ifthenelse{\boolean{cms@external}}{
    }{
\\
}
{$\lambda=0.625\pm0.021\,\text{(stat.)}\pm0.046\,\text{(syst.)}$} and
{$r=1.59\pm0.05\,\text{(stat.)}\pm0.19\,\text{(syst.)}\,\text{fm}$}  at $0.9$\TeV;
{$\lambda=0.663\pm0.073\,\text{(stat.)}\pm0.048\,\text{(syst.)}$}  and
{$r=1.99\pm0.18\,\text{(stat.)}\pm0.24\,\text{(syst.)}\,\text{fm}$} at $2.36$\TeV.
%

\begin{figure}[!t]
\ifthenelse{\boolean{cms@external}}{
    }{
    \begin{center}
    }
\includegraphics[height=0.48\textwidth,angle=90, trim = 0mm 16mm 10mm 0mm, clip]{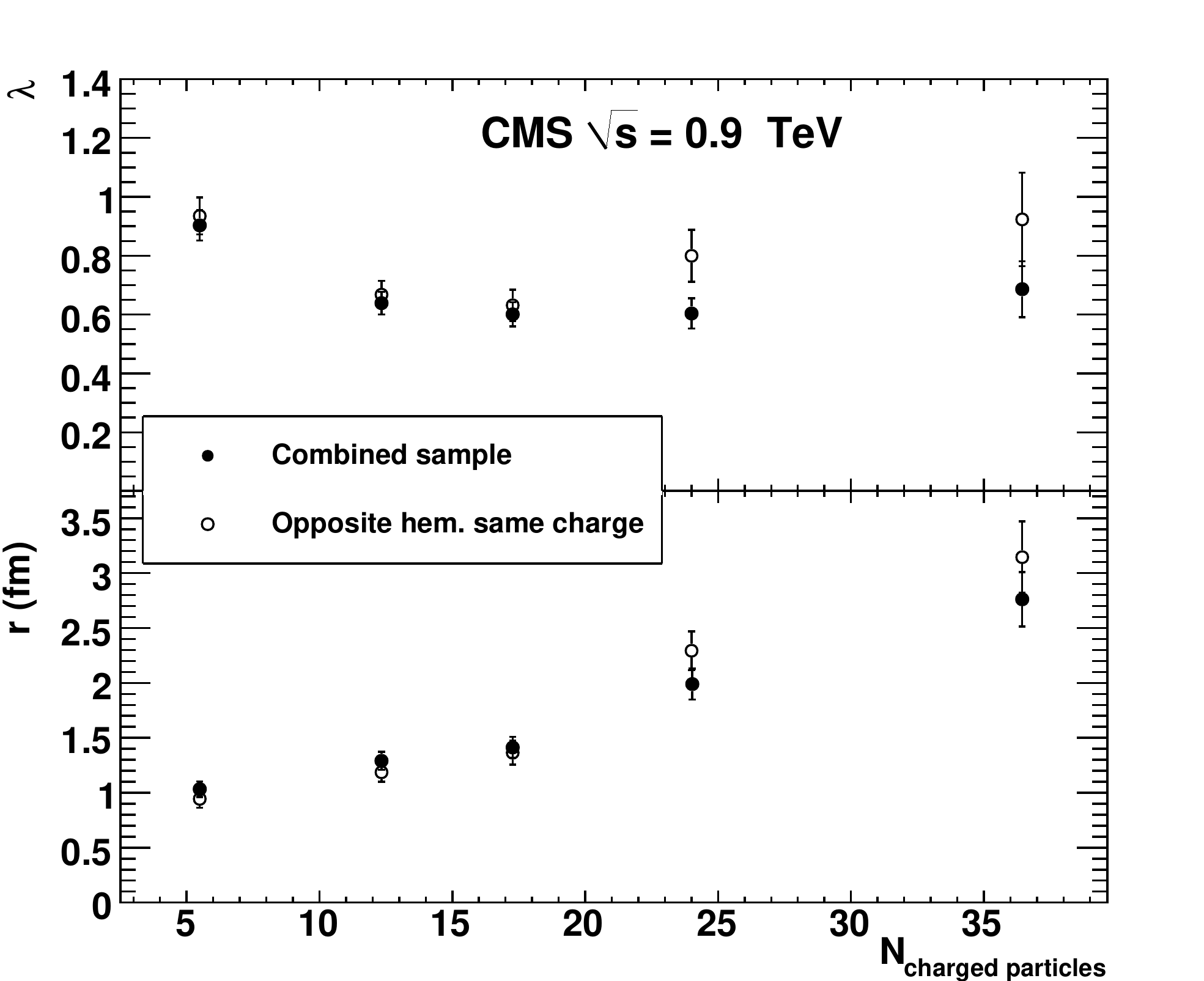}
        \caption{Values of the $\lambda$ (top) and $r$ (bottom) parameters as a
            function of the charged-particle multiplicity in the event for
		combined (dots) and opposite-hemisphere, same-charge (open
                circles) reference samples, at 0.9\TeV. The errors shown are statistical
            only. The points are placed on the horizontal scale at the average
            of the multiplicity distribution in the corresponding bin.}
        \label{f:vs_n}
\ifthenelse{\boolean{cms@external}}{
}{
\end{center}
}
\end{figure}
The possible dependence of the BEC signal on various track and event observables has been studied.
A significant dependence of $r$ on the charged-particle multiplicity in the event is observed
for all reference samples. Here, the only mixed-event reference sample used is the one constructed by combining charged particles from events in the same
multiplicity range.
The fit parameters for the combined reference sample are given in Table~\ref{t:rvsnch} and shown in Fig.~\ref{f:vs_n}
as a function of the track multiplicity for the $0.9\TeV$ data.
As an example,  the results for the opposite-hemisphere,
same-charge reference sample are also shown in Fig.~\ref{f:vs_n}.
The systematic errors on $\lambda$ and $r$ in each multiplicity bin are taken as the r.m.s.\ spread of the
results obtained with
the various reference samples.
\begin{table}[hbt]
\ifthenelse{\boolean{cms@external}}{
    }{
    \begin{center}
    }
\caption{Results of the fits to the double ratio ${\cal R}^{\text{comb}}$
for the combined reference samples, using the parameterization of Eq.~(\ref{eq:exp}) with the exponential form,
as a function of the charged-particle multiplicity in the event, for $0.9$\TeV data.
Errors are statistical only, except for $\lambda$ and $r$ where
statistical (first error) and systematic uncertainties (second error) are given.}
\label{t:rvsnch}
{
\ifthenelse{\boolean{cms@external}}{
\scriptsize
    }{
}
\begin{tabular}{c|c|c|c|c|c}
\hline
\hline
Mult. &       $p$ val.    &         $C$ &$\lambda$     & $r$ (fm)       & $\delta$ ($10^{-3}$  \\
range & (\%) & & & &  $\GeVinv$) \\
\hline
  2--9          & $97$  & $0.90$$\pm$$0.01$ & $0.89$$\pm$$0.05$$\pm$$0.20$ & $1.00$$\pm$$0.07$$\pm$$0.05$   & $72$$\pm$$12 $  \\
10--14        & $38$  & $0.97$$\pm$$0.01$ & $0.64$$\pm$$0.04$$\pm$$0.09$ & $1.28$$\pm$$0.08$$\pm$$0.09$   & $18$$\pm$$\,\,\,5 $  \\
15--19        & $27$  & $0.96$$\pm$$0.01$ & $0.60$$\pm$$0.04$$\pm$$0.10$ & $1.40$$\pm$$0.10$$\pm$$0.05$   & $28$$\pm$$\,\,\,5 $  \\
20--29        & $24$  & $0.99$$\pm$$0.01$ & $0.59$$\pm$$0.05$$\pm$$0.17$ & $1.98$$\pm$$0.14$$\pm$$0.45$   & $13$$\pm$$\,\,\,3 $  \\
30--79        & $28$  & $1.00$$\pm$$0.01$ & $0.69$$\pm$$0.09$$\pm$$0.17$ & $2.76$$\pm$$0.25$$\pm$$0.44$   & $10$$\pm$$\,\,\,3 $  \\
\hline
\end{tabular}
}
\ifthenelse{\boolean{cms@external}}{
    }{
\end{center}
}
\end{table}
Due to the limited sample size of the $2.36$\TeV data only two multiplicity bins are considered,
one for multiplicities smaller than 20 tracks, the other for multiplicities between 20 and 60 tracks.
The values measured for the parameters with the combined reference samples are
\mbox{$\lambda=0.65\pm0.08$} and \mbox{$\lambda=0.85\pm0.17$},
and \mbox{$r=1.19\pm 0.17$~fm} and \mbox{$r=2.85\pm 0.38$~fm}
for these two multiplicity bins, where the errors are statistical only.
For comparison, the values obtained for the same multiplicity bins at
\mbox{$0.9$\TeV} are
\mbox{$\lambda=0.65\pm0.02$} and \mbox{$\lambda=0.63\pm0.05$},  and
{$r=1.25\pm0.05$~fm} and \mbox{$r=2.27\pm0.12$~fm},
respectively. These measurements are consistent within errors.
The dependence of $r$ on multiplicity was already observed in previous
measurements as discussed in detail in~\cite{revbec}.

In summary,
Bose--Einstein correlations have been measured
for the first time at the LHC by the
CMS experiment in $pp$ collisions at $0.9$ and $2.36$\TeV center-of-mass energies.
Several reference samples were used to extract the signal. For all of them an
exponential shape fits the data significantly better than a Gaussian shape.
An increase of the effective size of the emission region with charged-particle multiplicity in the event
has been observed.

\ifthenelse{\boolean{cms@external}}{
}{
\section*{Acknowledgements \label{s:ack}}
}
We wish to congratulate our colleagues in the CERN accelerator departments for
the excellent performance of the LHC machine. We thank the technical and
administrative staff at CERN and other CMS institutes, and acknowledge support
from: FMSR (Austria); FNRS and FWO (Belgium); CNPq, CAPES, FAPERJ, and FAPESP
(Brazil); MES (Bulgaria); CERN; CAS, MoST, and NSFC (China); COLCIENCIAS
(Colombia); MSES (Croatia); RPF (Cyprus); Academy of Sciences and NICPB
(Estonia); Academy of Finland, ME, and HIP (Finland); CEA and CNRS/IN2P3
(France); BMBF, DFG, and HGF (Germany); GSRT (Greece); OTKA and NKTH (Hungary);
DAE and DST (India); IPM (Iran); SFI (Ireland); INFN (Italy); NRF (Korea); LAS
(Lithuania); CINVESTAV, CONACYT, SEP, and UASLP-FAI (Mexico); PAEC (Pakistan);
SCSR (Poland); FCT (Portugal); JINR (Armenia, Belarus, Georgia, Ukraine,
    Uzbekistan); MST and MAE (Russia); MSTDS (Serbia); MICINN and CPAN (Spain);
Swiss Funding Agencies (Switzerland); NSC (Taipei); TUBITAK and TAEK (Turkey);
STFC (United Kingdom); DOE and NSF (USA).

\bibliography{auto_generated}   

\cleardoublepage\appendix\section{The CMS Collaboration \label{app:collab}}\begin{sloppypar}\hyphenpenalty=5000\widowpenalty=500\clubpenalty=5000\textbf{Yerevan Physics Institute,  Yerevan,  Armenia}\\*[0pt]
V.~Khachatryan, A.M.~Sirunyan, A.~Tumasyan
\vskip\cmsinstskip
\textbf{Institut f\"{u}r Hochenergiephysik der OeAW,  Wien,  Austria}\\*[0pt]
W.~Adam, T.~Bergauer, M.~Dragicevic, J.~Er\"{o}, C.~Fabjan, M.~Friedl, R.~Fr\"{u}hwirth, V.M.~Ghete, J.~Hammer\cmsAuthorMark{1}, S.~H\"{a}nsel, M.~Hoch, N.~H\"{o}rmann, J.~Hrubec, M.~Jeitler, G.~Kasieczka, W.~Kiesenhofer, M.~Krammer, D.~Liko, I.~Mikulec, M.~Pernicka, H.~Rohringer, R.~Sch\"{o}fbeck, J.~Strauss, A.~Taurok, F.~Teischinger, W.~Waltenberger, G.~Walzel, E.~Widl, C.-E.~Wulz
% ...
\vskip\cmsinstskip
% ...
1:~~Also at CERN, European Organization for Nuclear Research, Geneva, Switzerland\\
% ...

\end{sloppypar}
\end{document}